\documentclass[lettersize,journal]{IEEEtran}
\usepackage{amsmath,amsfonts,amssymb}
\usepackage{algorithmic}
\usepackage{algorithm}
\usepackage{array}
\usepackage{textcomp}
\usepackage{stfloats}
\usepackage{url}
\usepackage{multirow}
\usepackage{tabularx,booktabs} 
\usepackage{verbatim}
\usepackage{graphicx}
\usepackage{cite}
\usepackage{makecell}
\usepackage{url}
\usepackage{relsize} 
\usepackage{color}
\usepackage{graphicx}
\usepackage{float}
\usepackage{subfigure}
\usepackage{enumitem}
\usepackage{bbm}
\usepackage{amsmath}
\usepackage{stfloats}
\makeatletter
\renewcommand{\maketag@@@}[1]{\hbox{\m@th\normalsize\normalfont#1}}%
\makeatother
\usepackage{hyperref}  

\definecolor{b}{rgb}{0,0,0} 

\begin{document}

\title{Connectivity Analysis of LoRaWAN-Based Non-Terrestrial Networks for Subterranean mMTC}

\author{Kaiqiang~Lin,~\IEEEmembership{Member,~IEEE} and
         Mohamed-Slim Alouini,~\IEEEmembership{Fellow,~IEEE}

\thanks{This work was supported in part by ENERCOMP Technology Consortium. (\textit{Corresponding author: Kaiqiang Lin})}

\thanks{K. Lin and M.-S. Alouini are with the Division of Computer, Electrical and Mathematical Sciences and Engineering, King Abdullah University of Science and Technology, Saudi Arabia (E-mail: kaiqiang.lin@kaust.edu.sa; slim.alouini@kaust.edu.sa).}

}



\maketitle

\begin{abstract}
Wireless underground sensor networks (WUSNs) offer significant social and economic benefits by enabling the monitoring of subterranean entities. However, the communication reliability of WUSNs diminishes in harsh environments where terrestrial network infrastructure is either unavailable or unreliable. To address this challenge, we explore the feasibility of integrating buried massive machine-type communication (mMTC) sensors with non-terrestrial networks (NTNs), including unmanned aerial vehicles (UAVs), high-altitude platforms (HAPs), and low Earth orbit (LEO) satellites, to establish underground-to-NTN connectivity for various large-scale underground monitoring applications. To assess the effectiveness of underground-to-NTN connectivity, we develop a Monte Carlo simulator that incorporates a multi-layer underground attenuation model, the 3GPP empirical path loss model for various NTN platforms, and two LoRaWAN modulation schemes, i.e., LoRa and LoRa-frequency hopping spread spectrum (LR-FHSS). Our results evidence that LoRa SF7 is a strong candidate for short-range UAV communication in rural environments, while LR-FHSS modulation proves to be a promising option for HAP and LEO satellite platforms in massive WUSNs scenarios thanks to its adequate link budget and robustness to the interference. Finally, we demonstrate that the success probability of underground-to-NTN connectivity using LoRa and LR-FHSS is significantly affected by factors such as the monitoring environment, the number of devices, burial depth, and the soil's volumetric water content.
\end{abstract}

\begin{IEEEkeywords}
Wireless underground sensor networks (WUSNs), non-terrestrial networks (NTNs), LoRaWAN, underground-to-NTN connectivity, underground monitoring.
\end{IEEEkeywords}

\section{Introduction}
\IEEEPARstart{W}{ireless} underground sensor networks (WUSNs) enable the monitoring of various subterranean entities using wirelessly connected underground devices (UDs). Although WUSNs typically operate with low data rate requirements and can tolerate latency in data transmission, energy conservation remains critical to ensuring long-term functionality and reliable connectivity~\cite{VuranWUSNsreview, LinMag}. Therefore, low-power wide-area network (LPWANs) has emerged as a promising solution for WUSNs, offering low power consumption, large communication range, and reliable connectivity in low-data-rate applications~\cite{LPWANoverview}. Recently, studies~\cite{LinModel, AnnaModel, GZWUSN} have explored the feasibility of enabling underground-to-aboveground communication using LoRaWAN technology, a promising LPWAN-grade technology designed for massive machine-type communication (mMTC). The initial results demonstrated that LoRaWAN-based WUSNs hold significant potential to advance various applications, including but not limited to precision agriculture, underground infrastructure monitoring, and post-disaster rescue operations~\cite{XDModel, LinAdhoc, WUSNReview}.

However, LoRaWAN-based WUSNs face notable challenges in remote regions, disaster-stricken areas, and harsh environmental conditions where terrestrial network infrastructure is either unavailable or unreliable, leading to coverage gaps and diminished communication reliability. To address these challenges, non-terrestrial networks (NTNs) are emerging as a transformative solution for achieving ubiquitous connectivity in future wireless communication systems. NTNs utilize platforms such as unmanned aerial vehicles (UAVs), high altitude platforms (HAPs), and satellites to provide seamless connectivity in areas where terrestrial networks are compromised or unavailable~\cite{NTNReview, NTNLink}. With their broad coverage footprints, robust line-of-sight (LOS) communication capabilities, and flexible deployment options, NTN platforms have been extensively explored to enhance communication reliability and energy efficiency through optimizing task scheduling, resource allocation, and mobility management in resource-limited environments~\cite{ZiyeTMC, ZiyeTVT, ZiyeTcom}. Building on these advances, our work investigates how NTNs can complement LoRaWAN-based WUSNs by mitigating coverage gaps and improving communication reliability. Driven by this, we advocate for the integration of NTNs with LoRaWAN-based WUSNs, implying underground-to-NTN connectivity, to support mMTC in diverse subterranean monitoring scenarios. 

\subsection{Motivations}
To enable LoRaWAN-enabled WUSNs in hard-to-reach or post-disaster areas, the authors in~\cite{LinMag, LinDRL, LinAdhoc} proposed integrating LoRaWAN-based WUSNs with low-Earth-orbit (LEO) satellites, thereby establishing underground direct-to-satellite (U-DtS) connectivity. Despite the major technical progress of LoRaWAN-based WUSNs and U-DtS connectivity, these studies exhibits three major limitations: (i) the use of simplified single-medium propagation models, which overlook the impact of underground multilayer media commonly encountered in practical applications such as pipeline monitoring and post-disaster rescue; (ii) exclusive consideration of LEO satellites, without exploring a broader range of NTN platforms such as UAVs and HAPs, or evaluating communication reliability across various deployment environments (e.g., rural, urban, and dense urban); and (iii) the absence of a comprehensive evaluation of underground-to-NTN success probability using both LoRa and the newly introduced LoRa frequency-hopping spread spectrum (LR-FHSS) modulation schemes.

\subsection{Contributions}
To fill these gaps, this study investigates whether introduction of underground multi-layer media and various NTN platforms across different monitoring environments would allow underground-to-NTN connectivity and, if yes which modulation schemes (e.g., LoRa and LR-FHSS) can achieve the reliable underground-to-NTN connectivity with the highest success probability. To the best of our knowledge, this work may be the first of its kind to assess the feasibility and performance of underground-to-NTN connectivity involving UAVs, HAPs, and LEO satellites, and to identify the optimal LoRaWAN modulation scheme for achieving robust connectivity. Our specific contributions are summarized as follows:
\begin{itemize}
    \item We conceptualize an underground-to-NTN architecture for underground pipeline monitoring scenarios, which can be generalized to other subterranean mMTC applications. To accurately evaluate the feasibility of this architecture and determine the optimal modulation scheme that achieves the highest success probability of underground-to-NTN connectivity, we develop a Monte Carlo simulator incorporating: (i) an empirical path loss model that incorporates absorption in multilayered media, refraction losses in the interfaces between different layers, and a realistic 3GPP aboveground attenuation model derived from extensive field measurements across various NTN platforms and environmental conditions; and (ii) two LoRaWAN modulation schemes, i.e., LoRa and LR-FHSS, with the capture effect~\cite{CaptureTII}.
    
    \item We utilize the developed framework to analyze the impact of various factors, including the monitoring environment, the number of UDs, and underground parameters (e.g., burial depth and volumetric water content (VWC)), on the success probability of underground-to-NTN connectivity under two modulation schemes across different elevation angles. This aids in the selection of appropriate NTN platforms and LoRaWAN modulation schemes to meet the communication reliability requirements of subterranean mMTC applications in diverse monitoring scenarios.

    \item Our numerical results indicate that LR-FHSS enables more reliable connectivity between massive UDs and HAPs/LEO satellites, offering greater robustness against co-channel interference compared to LoRa. Meanwhile, LoRa SF7 proves to be an effective wireless solution for short-range UAV communication. Additionally, LR-FHSS achieves a high success probability for underground-to-NTN connectivity in rural environments. However, establishing robust connectivity for HAP and LEO satellite platforms at low elevation angles remains challenging in urban and dense urban environments with harsh underground conditions (e.g., high VWC and large burial depths). To promote technical understanding and enable the replication of our results, the underground-to-NTN simulator is publicly available on~\cite{SimCode}.
\end{itemize}

The remainder of this article is structured as follows. Section~\ref{RelateWorkSec} provides an overview of WUSNs, LPWANs, and NTNs technologies. Section~\ref{SysModSec} describes the system model. Section~\ref{SimResSec} presents and discusses the numerical results for various NTN platforms using two LoRaWAN modulation schemes. Finally, Section~\ref{ConclusionSec} concludes the article and outlines potential directions for future research.

\section{Related works} \label{RelateWorkSec}
\subsection{LPWAN-based WUSNs Technologies}
Two network topology approaches used in WUSNs are ad hoc and centralized networks~\cite{WUSNReview}. In an ad hoc network topology, UDs transmit sensor data to aboveground gateways via multi-hop configurations. However, the communication distance between UDs is typically limited to around $10$~m, which significantly restricts the scalability of ad hoc networks for large-scale WUSN applications~\cite{U2URes, WUSNTII}. In contrast, the centralized network topology allows UDs to communicate directly with aboveground gateways without the need for repeaters, thereby reducing the routing steps that typically result in high energy consumption. Therefore, from the perspective of connectivity reliability and energy efficiency, centralized network topologies, such as LPWANs, are preferred for WUSNs.

Currently, LoRaWAN, NB-IoT, and Sigfox dominate the LPWAN market. Among these, NB-IoT and Sigfox are operated by dedicated mobile network operators, which limits their deployment flexibility and often makes them infeasible in remote or post-disaster areas. In contrast, LoRaWAN, based on the LoRa physical layer patented by Semtech~\cite{LoRaWANreview}, allows users to independently set up and manage private networks without relying on specific network operators, enabling greater implementation flexibility for WUSNs. Recent studies have experimentally investigated the potential of LoRa modulation, underlying LoRaWAN, for WUSNs. Specifically, the experimental results in~\cite{LinModel, LoRaField} show that the underground-to-aboveground communication range can exceed $50$~m at a burial depth of $0.4$~m when utilizing the LoRaWAN technology. Herein, the key transmission obstacle is the substantial losses in the subterranean wireless channel, which is mainly affected by the soil properties, VWC, burial depth, UD’s operating frequency, multi-path fading effects caused by gravel and roots, etc~\cite{SunModel, XDModel}. In more complex underground monitoring situations, such as pipeline monitoring, additional attenuation arises from the absorption in different media (e.g., gas, water, pipeline, asphalt) and refraction losses in media interfaces~\cite{PMWu, PMLei}.

Due to severe attenuation in underground portions, scarce network resources, and economic constraints, significant challenges remain in LoRaWAN-based WUSNs for large-scale underground monitoring. To further extend the network coverage and enable subterranean mMTC in remote and post-disaster areas, since a dense deployment of network infrastructure is economically infeasible, one viable solution is to increase the coverage of the gateway by deploying it on NTN platforms.

\subsection{NTN  Systems}
NTN systems refer to networks that operate through an air/spaceborne vehicle (e.g., UAVs, HAPs, and satellites) for communication~\cite{NTNMag, NTNReview}. Regarding satellites communication, this study focuses on LEO satellite platforms, which offer lower launch costs, reduced signal attenuation, and shorter propagation delays compared to medium-Earth-orbit and geostationary satellites.

\subsubsection{UAVs} They fly at low altitudes (e.g., a few hundred meters), offering the smallest coverage area but also the lowest propagation delays and losses compared to HAPs and LEO satellites. This low-altitude operation enables UAVs to provide flexible, on-demand wireless connectivity, particularly in areas where fixed terrestrial infrastructure is unavailable, such as during natural disasters or in remote locations. Thus, UAVs are more cost-effective than always-on terrestrial network systems. However, they require significant propulsion energy for sustained flight, limiting their ability to support long-term underground monitoring. Moreover, UAVs are highly weather-sensitive, with performance affected by adverse conditions such as wind or rain.

\subsubsection{HAPs} They operate at altitudes around $20$~km, providing wide coverage and flexibility for large geographic areas, often spanning hundreds of kilometers. Their high-altitude positioning can provide greater stability in flight compared to UAVs. Therefore, HAPs are well-suited for applications such as urban and suburban underground infrastructure monitoring, disaster recovery, and providing broadband internet to remote areas. However, they face challenges related to maintenance, requiring periodic refueling or servicing, and stabilization, particularly under adverse weather conditions.

\subsubsection{LEO satellites} They orbit the Earth at altitudes between $500$ and $2000$~km, offering the largest coverage footprint compared to UAVs and HAPs. They provide all-weather, $24/7$, and global connectivity, making them particularly valuable for continuous communication over large geographic areas, including both urban and remote locations. However, the launch and maintenance costs of LEO satellites are significantly higher than those of UAVs and HAPs. Meanwhile, strong attenuation in soil and the hundreds to thousands of kilometers distance between UDs and LEO satellites result in high path loss. Furthermore, the Doppler effect caused by the satellite's mobility induces frequency shifts in the radio signal, negatively affecting packet delivery ratio and energy efficiency~\cite{AsadDop}.

Table~\ref{ProConTab} summarizes the use cases, pros, and cons of these NTN platforms. The selection of the most suitable NTN platform should be guided by the specific requirements of subterranean mMTC applications. This highlights the need to evaluate the uplink success probability between UDs and various NTN platforms. Despite several studies have investigated the feasibility and energy efficiency of LoRaWAN technology for underground-to-satellite connectivity~\cite{LinDRL, LinMag, LinAdhoc}, research on underground-to-UAV and underground-to-HAP communication remains limited. Moreover, the previous studies on underground-to-satellite connectivity have overlooked the impact of underground multilayer media and monitoring environments. To address these gaps, this work develops a simulator to evaluate underground-to-NTN connectivity, incorporating a multilayer-media channel model and considering various NTN platforms across both rural and urban environments.

\begin{table}[!t]
\caption{Pros and Cons of Various NTN Platforms for Subterranean mMTC Applications}
\centering
\renewcommand{\arraystretch}{1.2}
\label{ProConTab}
\begin{tabular}{|m{0.06\textwidth}<{\raggedright}|m{0.11\textwidth}<{\centering}|m{0.11\textwidth}<{\centering}|m{0.11\textwidth}<{\centering}|}
\hline
\textbf{Platform} &\textbf{Scenario} &\textbf{Pros} &\textbf{Cons} \\
\hline
\textbf{UAV} &Short-range monitoring in remote/post-disaster areas  &Low path loss, flexible, and cost-effective &Limited coverage, short battery life, and weather-sensitive \\
\hline
\textbf{HAP} &Medium-range monitoring in urban/suburban areas &Great balance between cost, coverage, flexibility, and stability  &Maintenance complexity and weather vulnerability\\
\hline
\textbf{LEO Satellite} &Large-scale monitoring in urban and remote areas &All-weather, 24/7, and seamless connectivity &High deployment costs, severe path loss, and Doppler shifts \\
\hline
\end{tabular}
\end{table}

\subsection{LoRaWAN Modulation Schemes}
\subsubsection{LoRa}
LoRa modulation is a derivative of chirp spread spectrum technology utilized at the physical layer in LoRaWAN. It allows one to define the appropriate trade-offs between radio coverage and energy consumption by offering tunable physical layer parameters, such as spreading factor (SF), bandwidth (BW), and coding rate (CR). Despite its simplicity and widespread adoption, LoRa modulation suffers from limitations in network capacity and collision robustness due to its reliance on an Aloha-like medium access protocol without collision prevention mechanisms~\cite{LoRaWCL, LoRaTutorial}. To address the scalability and reliability challenges of conventional LoRa modulation, the LoRa Alliance has introduced LR-FHSS data rates (DRs) into the LoRaWAN protocol, aiming to support massive connectivity in NTN scenarios~\cite{LoRaStandard, FHSSMag}.

\subsubsection{LR-FHSS}
Unlike LoRa modulation, the novel LR-FHSS scheme employs frequency hopping  combined with Gaussian minimum shift keying modulation, which enhances robustness against co-channel interference. Currently, LR-FHSS supports only uplink communication, i.e., from UDs to the gateway. Note that LR-FHSS can coexist with LoRa modulation without requiring modifications to the network architecture, and modulation switching can be efficiently managed via a single adaptive data rate command from the network server.  

Previous studies have investigated the network scalability of LR-FHSS in direct-to-satellite (DtS) connectivity. Specifically, an analytical model was proposed in~\cite{AsadFHSS} to assess the feasibility of LR-FHSS for DtS scenarios. In~\cite{FHSSDtS1}, this model was enhanced by incorporating factors such as channel fading, noise, and the capture effect. The authors in~\cite{LinMag} extended LR-FHSS-enabled DtS connectivity into subterranean domains. The simulation results demonstrated that LR-FHSS significantly outperforms LoRa modulation in terms of network capacity and can provide reliable connectivity for smart agriculture in remote areas.

Despite significant technical advancements in LR-FHSS-enabled DtS connectivity, it remains unclear whether the introduction of absorption in multi-layer media and various NTN platforms across different monitoring environments would support underground-to-NTN connectivity and affect the superiority of LR-FHSS over LoRa. To answer this question, we first present an empirical path loss model for various NTN platforms, followed by an illustration of the implementation process for two LoRaWAN modulation schemes. Finally, we generate numerical results to assess the success probability of underground-to-NTN connectivity using LoRa and LR-FHSS.

\section{System Model} \label{SysModSec}
\begin{table*}[t]
\caption{\textbf{List of Symbols}}
\label{Symbolstab}
\centering
\renewcommand{\arraystretch}{1.35}
\begin{tabular}{|p{0.26\textwidth}|p{0.65\textwidth}|}
\hline
\textbf{Symbol} & \textbf{Definition} \\
\hline
$PL_u$ & Underground path loss \\
\hline
$L_m$ & Attenuation in underground media \\
\hline
$L_r$ & Refraction loss in medium interfaces \\
\hline
$L_f$ & Multipath fading attenuation in underground soils\\
\hline
$\alpha_k$ & Attenuation constant of the $k$-th medium \\
\hline
$l_k$, $d_k$ & Propagation path and thickness in the $k$-th medium \\
\hline
$z_k$ & Electromagnetic impedance of the $k$-th medium \\
\hline
$\mu_0$, $\varepsilon_0$ & Free-space permeability and permittivity \\
\hline
$\varepsilon_k$ & Real part of the complex dielectric constant of the $k$-th medium \\
\hline
$PL_{g2u}^{rural}$, $PL_{g2u}^{urban}$ & Path loss for the ground-to-UAV link in rural and urban environments \\
\hline
$PL_{g2h}^y$, $PL_{g2s}^y$ $\left(y\in\{\text{LOS}, \text{NLOS}\}\right)$ & Path loss for the ground-to-HAP/satellite links under LOS and non-LOS (NLOS) conditions \\
\hline
$L_{sf}^y$, $L_{cl}^y$ $\left(y\in\{\text{LOS}, \text{NLOS}\}\right)$ & Shadow fading and clutter losses for the ground-to-HAP/satellite links under LOS and NLOS conditions \\
\hline
$\delta^{LOS}$, $\delta^{NLOS}$ & Standard deviations of shadow fading $L_{sf}^{LOS}$ and $L_{sf}^{NLOS}$ \\
\hline
$X_r$, $X_u$ & Shadow fading for the ground-to-UAV link in rural and urban environments \\
\hline
$\delta_r$, $\delta_u$ & Standard deviations of $X_r$ and $X_u$ \\
\hline
$H_{uav}$, $H_{hap}$, $H_{leo}$ & Altitudes of the UAV, HAP, and LEO satellite \\
\hline
$d_{g2u}$, $d_{g2h}$, $d_{g2s}$ & Distances from the ground to the UAV, HAP, and LEO satellite \\
\hline
$f$ & Carrier frequency in Hz \\
\hline
$c$ & Speed of light \\
\hline
$\theta$ & Elevation angle of NTN platforms \\
\hline
$R_E$ & Earth radius \\
\hline
$P_r^{uav}$, $P_r^{hap}$, $P_r^{leo}$ & Received power at the UAV, HAP, and LEO satellite \\
\hline
$P_{ut}$, $G_{ut}$ & Transmit power and antenna gain of the UDs \\
\hline
$G_r^{uav}$, $G_r^{hap}$, $G_r^{leo}$ & Antenna gains of gateways on the UAV, HAP, and LEO satellite \\
\hline
$P_{LOS}$ & LOS probability for the ground-to-HAP/satellite links \\
\hline
$D_{snr}$ & Demodulation SNR threshold \\
\hline
$d_{i}$ &  Distance between the $i$-th interfering UD and the NTN platform. \\
\hline
$P_{snr}^q$, $P_{sir}^q$, $P_s^q$ $\left(q\in\{\text{uav}, \text{hap}, \text{leo}\}\right)$ & Probabilities of successful packet decoding without interference, decoding via capture effect, and successful packet delivery \\
\hline
$\sigma_w$ & Variance of additive white Gaussian noise \\
\hline
$NF$ & Noise figure of the LoRaWAN gateway \\
\hline
$B$ & Signal bandwidth \\
\hline
$\gamma$ & Signal-to-interference ratio threshold \\
\hline
$N$ & Number of UDs \\
\hline
$\lambda$ & Packet generation rate per UD \\
\hline
$N_h$, $t_h$, $\lambda_h$, $A_h$, $P_h$ & Number, length, arrival rate, average count, and reception probability of header replicas in LR-FHSS \\
\hline
$N_f$, $t_f$, $\lambda_f$, $A_f$, $P_f$ & Number, length, arrival rate, average count, and reception probability of payload fragments in LR-FHSS \\
\hline
$\zeta$ & Probability of successfully decoding a fragment \\
\hline
$N_l$ & Physical-layer payload size \\
\hline
$C$ & Number of channels within the occupied bandwidth \\
\hline
$P_{fhss}$ & Overall packet success probability in the LR-FHSS analytical model \\
\hline
\end{tabular}
\end{table*}

To investigate the feasibility and performance of integrating WUSNs and NTN for subterranean mMTC monitoring, we develop a Monte Carlo simulator to characterize the success probability of LoRaWAN-based underground-to-NTN connectivity across various NTN platforms in different environments. The following subsection provides details of the system model. The main symbols used in the system model are summarized in Table~\ref{Symbolstab}.

\subsection{Underground-to-NTN Architecture}
\begin{figure}[!t]
    \centering
    \includegraphics[width=3.45in]{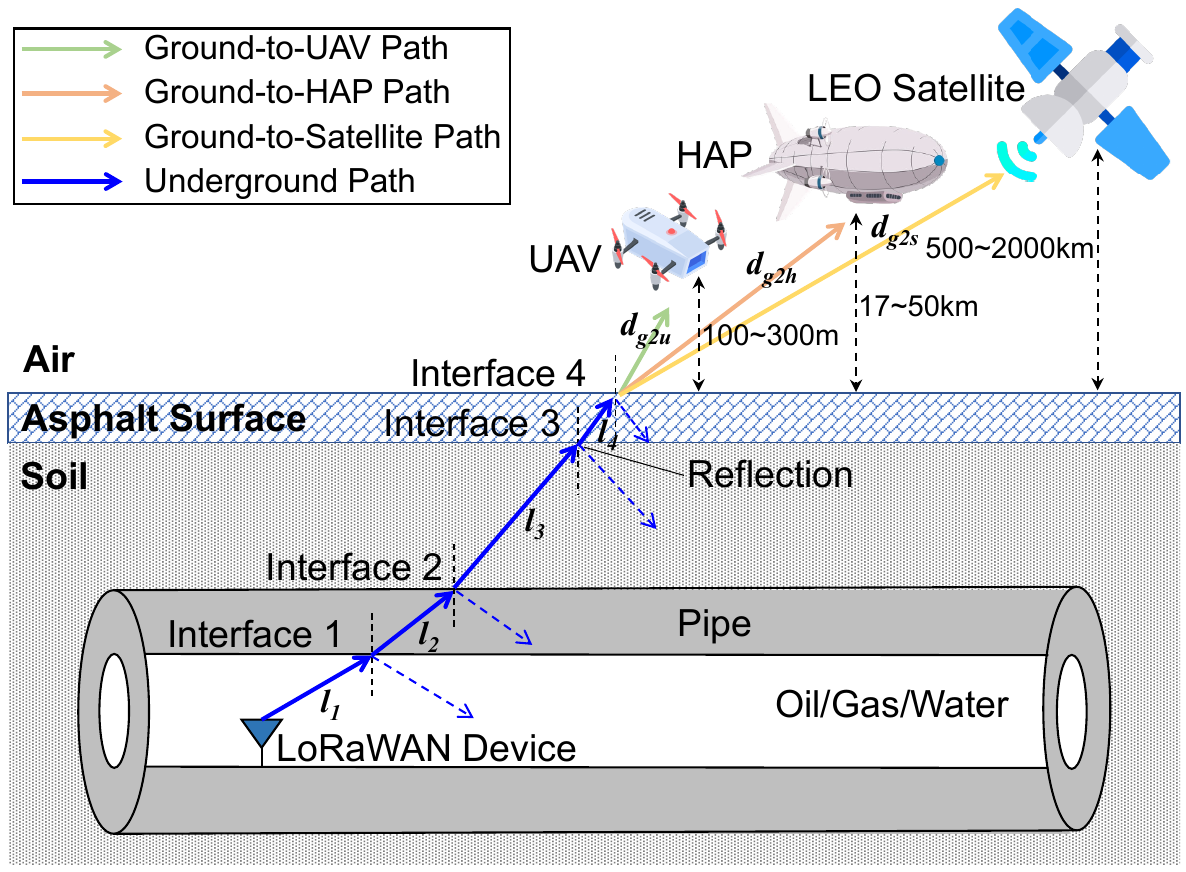}
    \caption{Radio propagation path of the underground-to-NTN communication channel in underground pipeline monitoring scenarios.}
    \label{fig_UNTNPath}
\end{figure}

We consider a comprehensive underground-to-NTN architecture, using an underground pipeline monitoring system as a representative example. As depicted in Fig.~\ref{fig_UNTNPath}, UDs are installed either inside or outside the pipeline to monitor key metrics such as pressure and noise levels on the pipeline wall, water/oil flow velocity, and soil properties~\cite{PM1, LinAdhoc, PMLei}. The sensor data is transmitted to a network server via a gateway deployed on an NTN platform for data processing and decision-making, enabling real-time leakage detection. Accordingly, this work focuses on uplink communication for data collection, assuming no downlink communication. This study focuses on the underground direct-to-NTN connectivity, implying that UDs directly communicate with a single NTN platform (e.g., UAV, HAP, or LEO satellite) without relying on relays or considering inter-platform links between NTN platforms.

Unlike traditional over-the-air wireless sensor networks, underground-to-NTN communication involves a highly complex propagation environment. In gas transportation scenarios, radio signals traverse multiple media layers, including gas, plastic, soil, asphalt, and air, while encountering various medium interfaces. Consequently, signal attenuation is significantly affected by absorption within each medium, refraction in interfaces, and multi-path fading caused by factors such as roots and gravel in the soil, as well as reflections from aboveground objects like trees and buildings. This study focuses on developing an underground-to-NTN channel model tailored for gas transportation applications. Notably, the proposed architecture can be extended to other subterranean mMTC monitoring scenarios, such as smart agriculture and post-disaster rescue operations. 

\subsection{Channel Model}
The development of the channel model is a fundamental block for characterizing the success probability of underground-to-NTN connectivity.

\subsubsection{Underground Path Loss Model}
As illustrated in Fig.~\ref{fig_UNTNPath}, the signal travels through multiple layers of media and their interfaces to reach the ground. The underground path loss consists of attenuation due to absorption in each medium $L_m$, refraction loss in the medium interfaces $L_r$, and attenuation caused by multipath fading in the underground soil $L_f$. Therefore, the path loss in the underground portion is expressed as
\begin{equation}
    PL_u = L_m+L_r+L_f.
\end{equation}
Herein, the attenuation in different medium is calculated by~\cite{MultilayerCal, PMLei}
\begin{equation}
    L_{m} = \sum_{k=1}^{K} 8.69 \alpha_k l_k,
\end{equation}
where $l_k$ is the propagation path in the $k$-th medium, while $\alpha_k$ is the attenuation constant of the $k$-th medium. Note that the signal can penetrate through the interfaces only if the incident angle is small. Hence, in this study, we assume the signal propagates vertically through each medium\footnote{Assuming vertical signal propagation reflects typical UD deployments and provides a conservative estimate of attenuation, enabling tractable analysis despite possible angular deviations caused by terrain variations or antenna misalignment.}, i.e., $l_k = d_k$, where $d_k$ is the thickness of the $k$-th medium. Accordingly, the refraction loss in the interfaces is given by~\cite{IndexCal}
\begin{equation}
    L_r  = \sum_{k}^{K-1}10 \log_{10} \left(1-\left| \frac{z_k-z_{k+1}}{z_k+z_{k+1}}\right|^2\right),
\end{equation}
where $z_k = \sqrt{\frac{\mu_0}{\epsilon_0 \varepsilon_k}}$ is the electromagnetic impedance of the $k$-th medium, $\mu_{0}$ is the free-space permeability, $\varepsilon_{0}$ is the free space permittivity, and $\varepsilon_k$ is the real part of the complex dielectric constant for the $k$-th medium. Furthermore, $L_f$ varies with underground environmental conditions and should ideally be determined through field measurements; in this study, it is set to $27$~dB based on prior empirical observations~\cite{SunModel, UPipeModel, PMLei, LinModel}.

\subsubsection{Ground-to-UAV Path Loss Model}
In this study, we assume a hovering UAV scenario, in which the UAV remains stationary relative to the UDs. Under this assumption, the impact of mobility on the ground-to-UAV (G2U) link is negligible, allowing the channel to be approximated as quasi-static. This simplification facilitates tractable modeling of propagation loss and enables a quantitative evaluation of LoRaWAN performance\footnote{UAV mobility influences fast fading characteristics and communication reliability along flight trajectories; thus, developing a comprehensive and realistic channel modeling framework that accounts for mobility effects represents a valuable direction for future work.}. According to the realistic 3GPP channel model proposed in~\cite{3GPPUAV}, the path loss for the G2U link is expressed as~\cite[Table B-2]{3GPPUAV}
\begin{align}
    PL_{g2u}^{rural} = &\max\left[23.9-1.8 \log_{10}(H_{uav}), 20\right] \log_{10}(d_{g2u}) \nonumber \\
    &+20\log_{10}(\frac{40 \pi f}{3})-180+X_r,\\
    PL_{g2u}^{urban} = &-152+22\log_{10}(d_{g2u})+20\log_{10}(f)+X_u, 
\end{align}
where $H_{uav}$ is the altitude of the UAV, $d_{g2u}$ is the distance from the ground surface to the gateway on the UAV, $f$ is the carrier frequency in Hz, while $X_r$ and $X_u$ account for the shadow fading, which follows a zero-mean log-normal distribution with $\delta_r=4.2e^{-0.0046 H_{uav}}$~dB and $\delta_u=4.64e^{-0.0066 H_{uav}}$~dB for rural and urban environments, respectively. 

\subsubsection{Ground-to-HAP/LEO Satellite Path Loss Model}
The attenuation caused by the troposphere, ionosphere, atmospheric gases, fog, clouds, and rain droplets for LoRaWAN operating in the sub-GHz frequency bands is typically negligible~\cite{SubGPL}. Therefore, the path loss for ground-to-HAP (G2H) and ground-to-satellite (G2S) links can be modeled as the basic path loss model proposed in~\cite{3GPPHAPSate}, which accounts for the signal's free space propagation, clutter loss, and shadow fading. It is expressed as
\begin{equation}
    PL_{x}^{y} = L_{fs}+L_{sf}^y+L_{cl}^y,
\end{equation}
where $ x \in \{\text{g2h}, \text{g2s}\}$, $y \in \{\text{LOS}, \text{NLOS}\}$, while $L_{fs}$ is the free space path loss, which is given by 
\begin{equation}
    L_{fs} = 20\log_{10} (f) + 20 \log_{10}(d_x) + 20 \log_{10}\left(\frac{4 \pi}{c}\right),
\end{equation}
where $c$ is the speed of light, $d_x$ is the propagation distance of the G2H and G2S links, expressed as a function of HAP/LEO altitude, i.e., $H_{z}, z \in \{\text{hap}, \text{leo}\}$, and their elevation angle $\theta$ by following the Earth radius $R_E=6371$~km,
\begin{equation}
    d_{x} = \sqrt{R_E^2 \sin^2(\theta)+H_z+2 H_z R_E} - R_E \sin(\theta).
\end{equation}
The clutter loss, $L_{sf}^y$ denotes the shadow fading loss characterized by a random number generated by a zero-mean normal distribution with standard deviation $\delta^y$, while $L_{cl}^y$ represents the clutter loss caused by buildings and environmental objects. The values of $L_{sf}^y$ and $L_{cl}^y$ at reference elevation angles for rural, urban and dense urban environments can be respectively found in~\cite[Table 6.6.2-1 -- 6.6.2-3]{3GPPHAPSate}. Note that the clutter loss can be negligible under the full LOS condition, implying $L_{cl}^{LOS}=0$. 

\subsection{Network Coverage Model}
\subsubsection{Received Power at NTN Platforms}
Since the propagation condition between UDs and the UAV can be considered as full LOS in both rural and urban environments when the UAV's altitude exceeds $100$~m, the received power for the UAV platform can be expressed as~\cite{3GPPUAV}
\begin{equation}
P_r^{uav} = P_{ut}+G_{ut}+G_{r}^{uav}-PL_u-PL_{g2u},    
\end{equation}
where $P_{ut}$ is the transmit power of UDs, while $G_{ut}$ and $G_{r}^{uav}$ are the antenna gains of the UDs and the gateway deployed on the UAV, respectively.

The received power at the HAP and LEO satellite platforms, $P_r^{z}$, $z \in \{\text{hap}, \text{leo}\}$, is given by~\cite{NTNLink} 
\begin{align}
P_r^{z} = &P_{ut}+G_{ut}+G_{r}^{z}-PL_u-P_{LOS} PL_{x}^{LOS} \nonumber \\
&- (1-P_{LOS}) PL_{x}^{NLOS},   
\end{align}
where $G_{r}^{z}$ represents the antenna gain of the gateway on the HAP or LEO satellite, while $P_{LOS}$ denotes the LOS probability, which depends on the elevation angle $\theta$ and the monitoring environment. The values of $P_{LOS}$ are empirically determined and can be found in~\cite[Table 6.6.1-1]{3GPPHAPSate}.

\subsubsection{Success Probability under LoRaWAN}
LoRaWAN technology demonstrates inherent robustness against Doppler effects; therefore, we neglect Doppler-induced impairments for a more straightforward and tractable system model in the underground-to-LEO satellite connectivity\footnote{A rigorous assessment of Doppler effects on the success probability of LoRaWAN remains an open and promising direction for future research, particularly when LoRa modulation operates in higher frequency bands with a high SF such as SF12.}~\cite{AsadDop}. For illustration purposes, we assume a LEO satellite mega-constellation that provides continuous coverage to UDs, and we evaluate the success probability of the G2S link during the satellite’s visibility period. Based on these assumptions, the success probability model is detailed as follows.

In the absence of interference, the gateway on the NTN platforms successfully demodulates the uplink packet if the received signal-to-noise ratio (SNR) exceeds the specific SNR demodulation threshold $D_{snr}=\{-6, -15, -20, 4\}$~dB for LoRa SF7, SF10, SF12, and LR-FHSS, respectively. Consequently, the probability of successfully decoding the packet is given by   

\begin{equation}
    \label{P_SNR} P_{snr}^{q} = \mathbb{P}\left[\frac{P_r^{q}(d)}{\sigma_w^2} \geq D_{snr}\right],
\end{equation}
where $\sigma_w^2=-174+NF+10 \log_{10} (B)$~dBm is the variance of the additive white Gaussian noise, $NF = 6$~dB is the noise figure of the LoRaWAN gateway's design architecture\footnote{This fixed noise figure, which aligns with typical gateway specifications, offers a practical baseline for analysis and ensures consistency across simulations despite potential variations in real-world deployments~\cite{LoRaWANreview}.}, $B$ is the signal BW, while $P_r^{q} (d)$ denotes the received power at the UAV, HAP, and LEO satellite platforms at a given distance $d$, with $q \in \{\text{uav}, \text{hap}, \text{leo}\}$. 

The recent studies have demonstrated that the capture effect enables the gateway to successfully recover the target packet during packet collisions, if the signal-to-interference ratio (SIR) of the target packet exceeds the SIR threshold $\gamma=6$~dB~\cite{CaptureTII}. Therefore, the probability of successfully demodulating the target packet by leveraging the capture effect under same-channel interference can be expressed as
\begin{equation}
\label{P_SIR} P_{sir}^{q} =\mathbb{P}\left[\frac{P_r^{q}(d)}{\sum P_{r}^{q}(d_i)} \geq \gamma\right],
\end{equation}
where $d_i$ denotes the distance between the $i$-th interfering UD and the NTN platforms. We assume that the collisions of packets featuring different SFs are quasi-orthogonal, thus this study focuses on utilizing the capture effect model to characterize the effects of co-channel and co-SF interference on the success probability of underground-to-NTN connectivity~\cite{LoRaWCL, LinAdhoc}.

Accordingly, we define the network coverage as the overall probability of successful packet delivery, which is the product of $P_{snr}$ and $P_{sir}$, i.e.,
\begin{equation}
\label{P_S} P_{s}^{q} = P_{snr}^{q} P_{sir}^{q}.
\end{equation}

\subsection{LoRaWAN Modulation Scheme Implementation}
\begin{figure}[!t]
    \centering
    \includegraphics[width=3.45in]{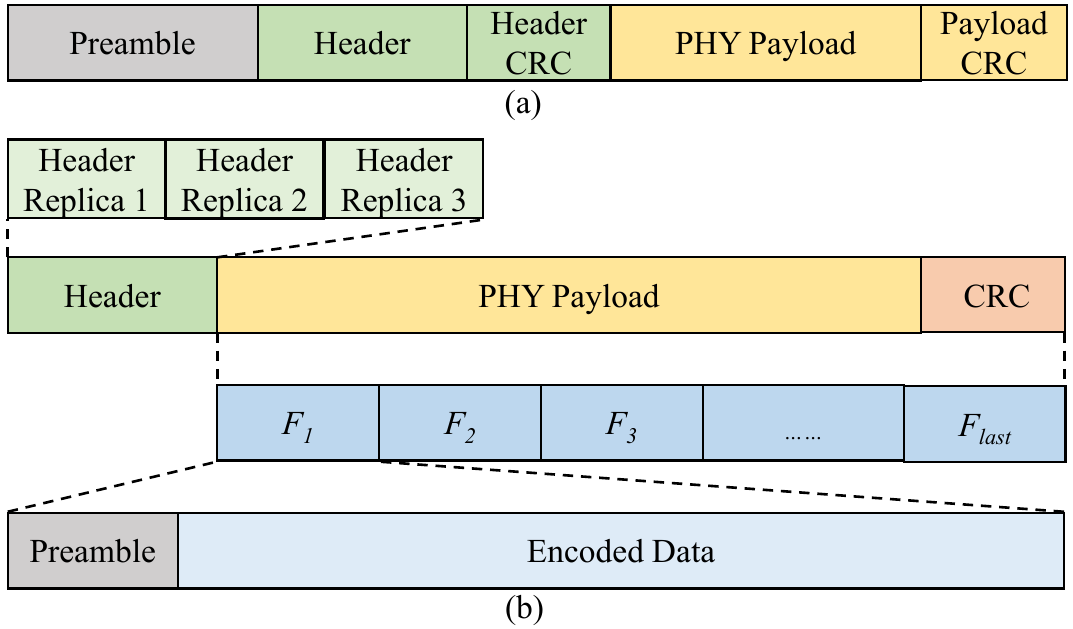}
    \caption{Packet structure of (a) LoRa and (b) LR-FHSS.}
    \label{fig_LoRaFHSSpacket}
\end{figure}

\begin{figure}[!t]
    \centering
    \includegraphics[width=3.45in]{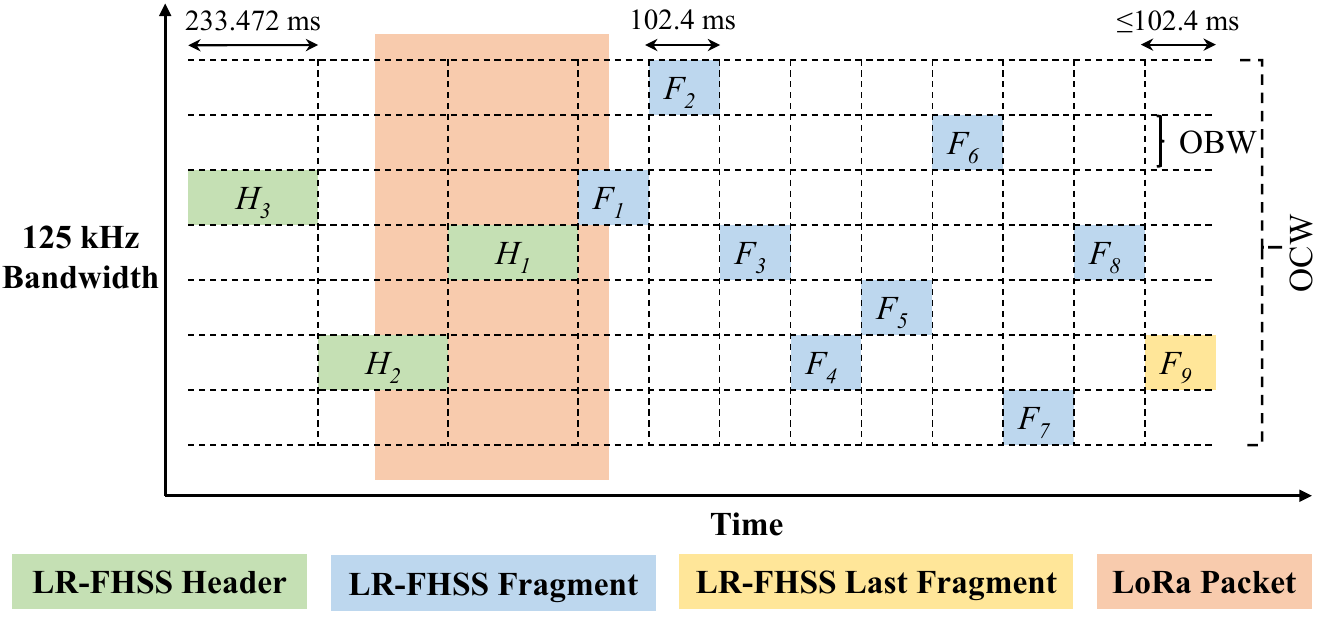}
    \caption{Transmission diagram for a single packet under LoRa and LR-FHSS modulation schemes, where LR-FHSS operates with DR8, featuring a code rate of $1/3$ and three header replicas.}
    \label{fig_LoRaFHSStrans}
\end{figure}

The primary difference between the LoRa and LR-FHSS modulation schemes lies in their packet structures and transmission procedures, as illustrated in Figs.~\ref{fig_LoRaFHSSpacket} and~\ref{fig_LoRaFHSStrans}, respectively. As shown in Fig.~\ref{fig_LoRaFHSSpacket}, a typical LoRa packet comprises three main components: a preamble, an optional header with a cyclic redundancy check (CRC), and a data payload also accompanied by a CRC. In contrast, LR-FHSS generates multiple header replicas and splits the payload into smaller fragments. As shown in Fig.~\ref{fig_LoRaFHSStrans}, LR-FHSS-based UDs transmit each packet fragment (i.e., the header or a payload fragment) over a randomly-selected occupied BW (OBW) frequency channel. Conversely, LoRa transmits the entire packet over the whole channel BW, resulting in a higher co-channel interference probability. The following sections present the details of the simulator on the packet reception procedures for both LoRa and LR-FHSS modulation schemes.

\subsubsection{LoRa Modulation}
\begin{figure}[!t]
    \centering
    \includegraphics[width=3.45in]{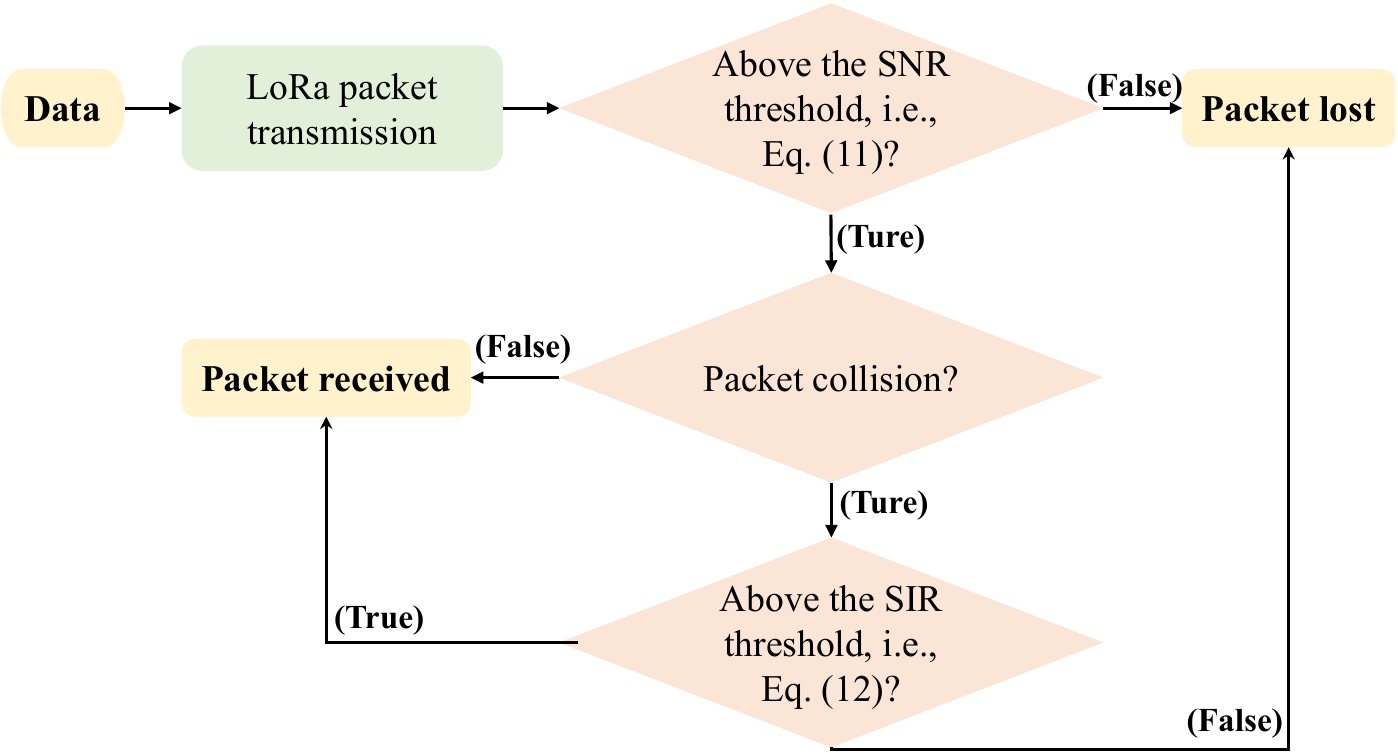}
    \caption{\textcolor{b}{Flow diagram for determining the successful reception of a target packet under the LoRa modulation scheme.}}
    \label{fig_LoRaPro}
\end{figure}

Under LoRa modulation, the process of determining whether a target packet is successfully received is depicted in Fig.~\ref{fig_LoRaPro}. First, the sensor data is converted into the LoRa packet format as described in~Fig.~\ref{fig_LoRaFHSSpacket}(a) and then randomly occupies the entire channel BW for uplink transmission as illustrated in Fig.~\ref{fig_LoRaFHSStrans}. The simulation model then evaluates whether the received packet's SNR exceeds the demodulation SNR threshold $D_{snr}$. If the SNR of the received packet exceeds $D_{snr}$, i.e., Eq.~\eqref{P_SNR}, and no packet collision occurs, the packet is deemed successfully received by the gateway. However, if multiple interfering packets with the same SF are transmitted simultaneously on the same channel as the target packet, leading to a packet collision, further evaluation is required to determine whether the target packet can still be recovered. If the capture effect condition is met, as defined in Eq.~\eqref{P_SIR}, the gateway can successfully demodulate the target packet. Otherwise, the packet reception is considered unsuccessful.

\subsubsection{LR-FHSS Modulation}
\begin{figure*}[!t]
    \centering
    \includegraphics[width=5in]{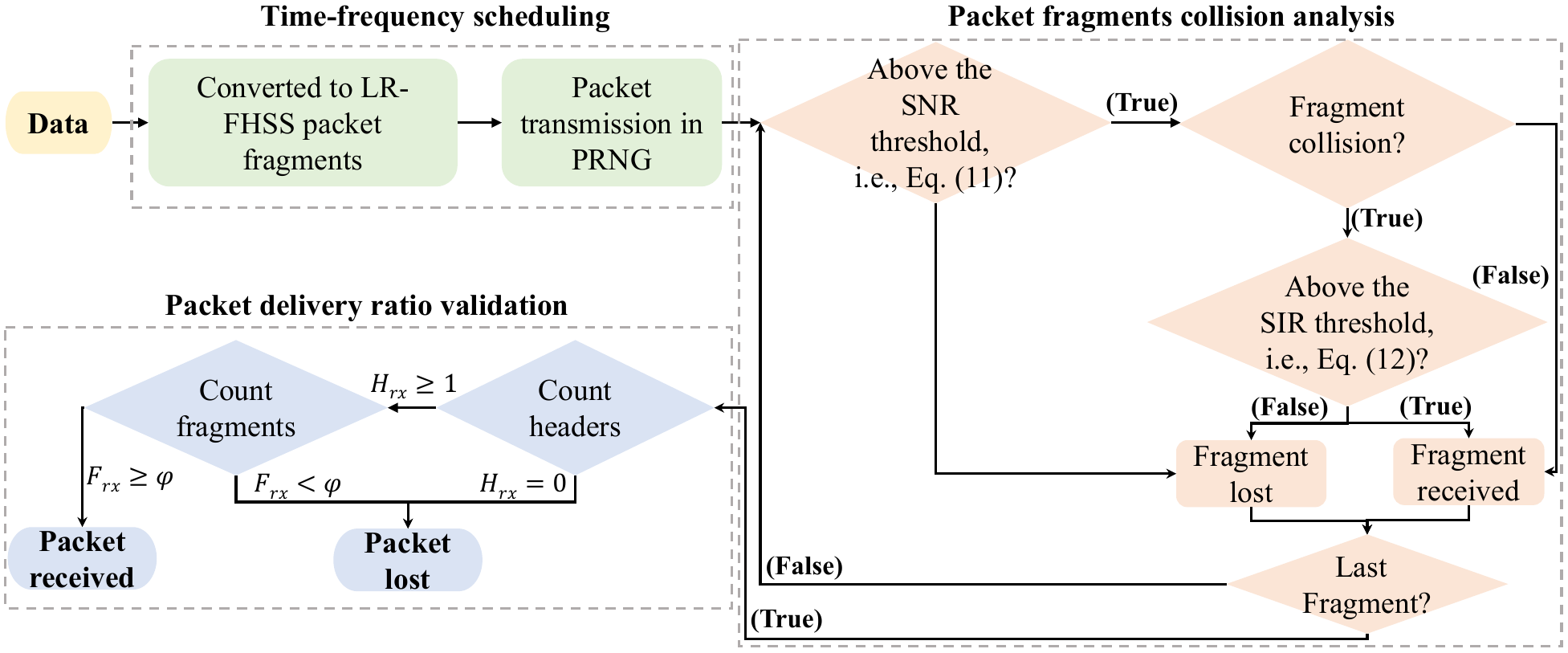}
    \caption{\textcolor{b}{Flow diagram for determining the successful reception of a target packet under the LR-FHSS modulation scheme.}}
    \label{fig_FHSSPro}
\end{figure*}
Unlike LoRa modulation, the novel LR-FHSS scheme exploits frequency hopping and offers high robustness against co-channel interference through increasing the number of physical channels, adding redundant physical headers, and lowering CR. We first present LR-FHSS analytical model without capture effects and then describe the procedure on determining the successful packet reception. 

In LR-FHSS, the frequency band is divided into operating channel width (OCW) channels, with BWs varying according to local frequency regulations. Each OCW channel is further subdivided into multiple physical frequency channels, termed OBW channels, each with a BW of $488$ Hz. For uplink communication, an LR-FHSS-enabled UD initiates transmission by sending $N_h$ replicas of a $t_h = 233.472$~ms-long header, using a new randomly-selected carrier frequency for each replica. Then, the $N_l$ bytes of physical-layer (PHY) payload and CRC are split into $N_f$ fragments, each with a maximum duration of $t_f = 102.4$~ms, and transmitted sequentially over OBW frequency channels selected by a pseudorandom number generator (PRNG). Successful uplink packet reception at the gateway requires decoding at least one header replica and either $1/3$ or $2/3$ of the payload fragments, depending on the CR. The key LR-FHSS parameters for different regions are summarized in Table~\ref{LRFHSSTab}. In this study, we focus on the DR8 configuration for LR-FHSS to evaluate its performance in comparison to LoRa. Our developed simulator can be adapted to other regions by modifying the LR-FHSS parameters.

\begin{table*}[ht]
\caption{Key Parameters of LR-FHSS DRs for ETSI and FCC regions}
\centering
\renewcommand{\arraystretch}{1.1}
\begin{tabular}{|c|cc|cccc|}
\hline
Region     & \multicolumn{2}{c|}{FCC (902$\sim$928~MHz)}                    & \multicolumn{4}{c|}{ETSI (863$\sim$870~MHz)} \\ \hline
LoRaWAN DRs     & \multicolumn{1}{c|}{DR5}       & DR6       & \multicolumn{1}{c|}{DR8}     & \multicolumn{1}{c|}{DR9}     & \multicolumn{1}{c|}{DR10}    & DR11    \\
OCW BW & \multicolumn{1}{c|}{1.523 MHz} & 1.523 MHz & \multicolumn{1}{c|}{137 kHz} & \multicolumn{1}{c|}{137 kHz} & \multicolumn{1}{c|}{336 kHz} & 336 kHz \\
OBW  channels & \multicolumn{1}{c|}{3120}      & 3120      & \multicolumn{1}{c|}{280}     & \multicolumn{1}{c|}{280}     & \multicolumn{1}{c|}{688}     & 688     \\
Grid separation & \multicolumn{1}{c|}{25.4 kHz}  & 25.4 kHz  & \multicolumn{1}{c|}{3.9 kHz} & \multicolumn{1}{c|}{3.9 kHz} & \multicolumn{1}{c|}{3.9 kHz} & 3.9 kHz \\
Number of header copies & \multicolumn{1}{c|}{3}         & 2         & \multicolumn{1}{c|}{3}       & \multicolumn{1}{c|}{2}       & \multicolumn{1}{c|}{3}       & 2       \\
CR    & \multicolumn{1}{c|}{1/3}       & 2/3       & \multicolumn{1}{c|}{1/3}     & \multicolumn{1}{c|}{2/3}     & \multicolumn{1}{c|}{1/3}     & 2/3     \\ \hline
\end{tabular}
\label{LRFHSSTab}
\end{table*}

Following the analytical model developed in~\cite{AsadFHSS}, we assume that each UD transmits, on average, one message every $1/\lambda$ units of time. Accordingly, the average inter-arrival times between consecutive header replicas and payload fragments are modeled as exponential random variables with rates $\lambda_h = N_h \lambda N$ and $\lambda_f = N_f \lambda N$, respectively, where $N$ is the total number of UDs. Each packet fragment is subject to a success probability $\zeta$, which is considered as independent and identically distributed (i.i.d.) across time, frequency, and space. Here, $\zeta$ implies the probability that the received SNR of a packet fragment exceeds the demodulation threshold in the absence of interference, i.e., $P_{snr}^q$ in Eq~\eqref{P_SNR}. Assuming that all payload fragments are of equal size, the average number of packet fragments arriving at the gateway during the header and payload transmission phases are respectively expressed as~\cite{AsadFHSS} 
\begin{align}
    A_h &= 2 \lambda_h  t_h + \lambda_f(t_h+t_f), \\
    A_f &= 2 \lambda_f  t_f + \lambda_h(t_h+t_f).
\end{align}
A packet is considered successfully received at the gateway when the following two independent conditions are simultaneously satisfied: (i) at least one of the $N_h$ header replicas is successful received, which occurs with probability given by
\begin{equation}
    P_h = 1 - \left(1-\zeta\left(1-\frac{1}{C}\right)^{A_h-1}\right)^{N_h}, \label{pheq}
\end{equation}
where $C$ denotes the number of channels within an OBW channel for frequency hopping, (ii) at least $\varphi$ payload fragments are correctly received, with the probability given by
\begin{equation}
    P_f = 1 - \sum_{k=0}^{\varphi - 1} \binom{N_f}{k} p_o^k (1 - p_o)^{N_f - k}, \label{pfeq}
\end{equation}
where $\binom{a}{b} = [a!/b!(a-b)!]$ represents the binomial coefficient, while $p_o = \zeta \left(1 - \frac{1}{C} \right)^{A_f - 1}$ is the probability of successfully receiving a single payload fragment. 

The overall packet success probability in LR-FHSS is obtained by combining Eqs.~\eqref{pheq} and~\eqref{pfeq}, and is given by
\begin{equation}
    P_{fhss} = P_h P_f. 
\end{equation}

To enable a fair comparison with LoRa modulation and evaluate LR-FHSS performance in complex interference scenarios that are difficult to analyze analytically, we developed a Monte Carlo simulation model to determine whether a target packet is successfully received at the gateway, accounting for the capture effect. The packet reception process in the simulator under LR-FHSS modulation is illustrated in Fig.~\ref{fig_FHSSPro}, and is divided into three stages.  

\color{black}
\begin{enumerate}
    \item Time-frequency scheduling. The target packet is formatted into $N_h$ replicas of the $233.472$~ms-long header and $N_f$ payload fragments with a time-on-air not exceeding $102.4$~ms, as shown in Fig.~\ref{fig_LoRaFHSSpacket}(b). The number of header replicas is determined by the DR setting. During a complete transmission cycle, each packet fragment is sent on a random OBW channel at an instantaneous time, following PRNG, as illustrated in Fig.\ref{fig_LoRaFHSStrans}.
    
    \item Packet fragments collision analysis. The simulator assesses whether the packet fragment meets the demodulation SNR threshold, as specified in Eq.~\eqref{P_SNR}. If the SNR threshold is not met, the packet fragment is considered to have failed reception. If the threshold is satisfied, the simulator proceeds to assess the occurrence of a collision, which is defined as an overlap in both time and frequency channels between two or more packet fragments. In the absence of a collision, the packet fragment is successfully received. However, if a collision is detected, the simulation model further checks whether the capture effect condition, as defined in Eq.~\eqref{P_SIR}, is satisfied. If the capture effect condition is met, the packet fragment is successfully demodulated; otherwise, it is considered lost. This process is repeated until all packet fragments have been processed.
    
    \item Packet delivery ratio validation. To successfully decode the target packet, the gateway must receive at least one header (i.e., $H_{rx} \geq 1$) and the number of received payload fragments should exceed the pre-defined reception threshold determined by the CR (i.e., $F_{rx} \geq \varphi$). If either condition is not met, the target packet is considered lost.
\end{enumerate}

\subsection{System Objective}
The primary objective of this work is to assess the feasibility of using LoRaWAN technology to establish reliable underground-to-NTN communication, thereby enabling large-scale WUSNs to operate in remote, inaccessible, or disaster-affected areas. To this end, we develop a simulator to identify the optimal LoRaWAN modulation scheme that maximizes the success probability of underground-to-NTN connectivity across various NTN platforms and elevation angles. Given the candidate modulation schemes $\mathbf{c} = \{\text{LoRa, LR-FHSS}\}$, the objective of maximizing the uplink success probability is formulated as follows:
\begin{subequations}\label{p1}
\begin{align} 
    ({\rm{P1}}):~{\mathop {\max} \limits_{\mathbf{c}}} ~&{P_s^q}\\  
    \text{s.t.}~ \label{c1_1} &10^{\circ} \leq \theta \leq 90^{\circ},
\end{align}
\end{subequations}
where $P_s^q$ is the overall probability of successful packet delivery, with $q \in \{\text{uav}, \text{hap}, \text{leo}\}$ indicating the selected NTN platform, i.e., UAVs, HAPs, or LEO satellites, respectively. Constraint~\eqref{c1_1} specifies that the elevation angle $\theta$ ranging from $10^{\circ}$ to $90^{\circ}$ is considered for all NTN platforms in this study. Note that while LoRa offers six modulation configurations, this study focuses on three representative ones, i.e., SF7, SF10 and SF12. 

To determine the optimal modulation scheme for practical underground-to-NTN deployments, the simulator first requires users to specify the NTN platform (e.g., UAVs, HAPs, or LEO satellites) and the corresponding elevation angle based on the target monitoring scenario. It then evaluates both LoRa and LR-FHSS modulation schemes to identify the configuration that maximizes the uplink success probability. Accordingly, in Section~\ref{SecResMS}, we present the success probability results of both modulation schemes across different elevation angles for each NTN platform in a rural environment. Since the performance of each modulation scheme is strongly influenced by factors such as the deployment environment (e.g., rural, urban, or dense urban), the number of UDs, the burial depth, and the soil's VWC, we provide a comprehensive analysis of how these parameters affect success probability and highlight the optimal modulation scheme for ensuring reliable underground-to-NTN connectivity in Sections~\ref{SecResEnv},~\ref{SecResNum}, and~\ref{SecResUnderground}.

\color{black}
\section{Simulation Results and Discussion}\label{SimResSec}
To analyze the feasibility and performance of underground-to-NTN connectivity using LoRa and LR-FHSS modulation schemes for subterranean mMTC applications in various environments (i.e., rural, urban, and dense urban), we conduct MATLAB-based Monte Carlo simulations via the specially developed simulator as detailed in Section~\ref{SysModSec}. Our proposed simulator is made accessible to the public via GitHub~\cite{SimCode}. 

\begin{table}[!t]
\caption{Configurations and Parameters for Monte Carlo Simulations}
\centering
\renewcommand{\arraystretch}{1.2}
\begin{tabular}{|m{0.28\textwidth}<{\raggedright}|m{0.15\textwidth}<{\centering}|}
\hline
\textbf{Parameters} & \textbf{Values} \\  
\hline
\multicolumn{2}{|l|}{\textbf{Operation Environments}} \\ 
\hline
Total number of UDs ($N$) & 50k \\
UDs' deployment & Uniform and random \\
Diameter of the pipeline ($d_1$) & 0.15~m \\
Thickness of the pipeline body ($d_2$) & 0.05~m \\
Burial depth (i.e., depth of soil) ($d_3$) & 0.6~m \\
Thickness of the asphalt layer ($d_4$) & 0.1~m \\
VWC ($m_v$) & 11.19\(\%\)~(\textit{in-situ}) \\
Clay ($m_c$) & 16.86\(\%\)~(\textit{in-situ}) \\
Dielectric constant of gas, plastic, soil, asphalt, and air ($\varepsilon_1, \varepsilon_2, \varepsilon_3, \varepsilon_4, \varepsilon_5$) & 1, 3, 5.8, 7, 1 \\
PHY payload ($N_l$) & 10~bytes \\
Report period ($T$) & 600~s \\
Traffic pattern & Periodic \\
\hline
\multicolumn{2}{|l|}{\textbf{Radio Configuration}} \\ 
\hline
Carrier frequency ($f$) & 433~MHz \\
Antenna gain of UDs ($G_{ut}$) & 2.15~dBi \\
Transmit power of UDs ($P_{ut}$) & 14~dBm \\
SIR threshold (\(\gamma\)) & 6~dB \\
\hline
\multicolumn{2}{|l|}{\textbf{LoRa Modulation}} \\ 
\hline
SF & 7, 10, 12 \\
LoRa BW & 125~kHz \\
Frequency channels & 80 \\
Receiver sensitivities of SF7, SF10, and SF12 & -123~dBm, -132~dBm, -137~dBm \\
\hline
\multicolumn{2}{|l|}{\textbf{LR-FHSS Modulation}} \\ 
\hline
OBW channels ($C$) & 280 \\
Header replicas ($N_h$) & 3 \\
CR & 1/3 \\
Header replica duration ($t_h$) & 233.472~ms \\
Fragment duration ($t_f$) & 102.4~ms \\
LR-FHSS OBW BW & 488~Hz \\
LR-FHSS receiver sensitivity & -137~dBm \\
\hline
\multicolumn{2}{|l|}{\textbf{NTN Configuration}} \\ 
\hline
Elevation angles ($\theta$) & \(10^\circ \leq \theta \leq 90^\circ\) \\
Orbital height of UAV, HAP, and LEO satellite ($H_{uav}$, $H_{hap}$, $H_{leo}$) & 100~m, 20~km, 550~km \\
Antenna gain of UAV, HAP, and LEO satellite ($G_r^{uav}$, $G_r^{hap}$, $G_r^{leo}$) & 2~dBi, 17~dBi, 35~dBi \\
\hline
\end{tabular}
\label{ConfigTab}
\end{table}

The presented results were obtained through $10^5$ Monte Carlo iterations. The key configurations and parameters of the system model are summarized in Table~\ref{ConfigTab}. Specifically, we consider that $N=50,000$ (i.e., 50k) UDs are installed inside the pipeline at the same depth (i.e., $d_3=0.6$~m) to monitor gas pipeline leakage in Saudi Arabia, while each UD uploads a $10$-byte PHY payload packet every $T=600$~s. The \textit{in-situ} soil properties, e.g., VWC and clay percentage, is obtained from~\cite{ArbiaStudyArea} to accurately capture the path loss in the underground soil. Note that these parameters (e.g., the number of UDs, burial depths, and VWC of soil) should be adjusted according to the practical requirements and regional conditions of specific underground pipeline monitoring scenarios, thus we analyze the impact of varying these parameters on the success probability in Sections~\ref{SecResNum} and~\ref{SecResUnderground}. The carrier frequency is set to $f=433$~MHz, in accordance with regional regulations in Saudi Arabia\footnote{Since the underground direct-to-NTN connectivity in this study operates within a single regulatory region using a fixed sub-GHz ISM frequency band (i.e., $433$~MHz), there is no need for costly frequency-switching mechanisms. Note that cross-platform coordination and frequency-switching protocols should be considered in scenarios involving inter-NTN communication and global-scale connectivity.}~\cite{LoRaStandard}. This setup implies: (i) for UDs, the maximum transmit power is $P_{ut} = 14$~dBm with a $G_{ut} = 2.15$~dBi-gain antenna; (ii) a maximum of $80$ channels can be used for uplink communication under LoRa modulation; (iii) under LR-FHSS modulation, we focus on the DR8 configuration, allowing each UD to leverage $280$ OBW channels with $N_h = 3$ header replicas and a coding rate of $\frac{1}{3}$. Furthermore, we envision deploying LoRaWAN gateways on various NTN platforms, including UAVs, HAPs, and LEO satellites, depending on the specific requirements of subterranean mMTC applications. Note that all UDs and gateways are configured with identical LoRaWAN settings.

\subsection{Effects of LoRaWAN Modulation Schemes} \label{SecResMS}
\begin{figure*}[!ht]
    \centering
    \includegraphics[width=6.9in]{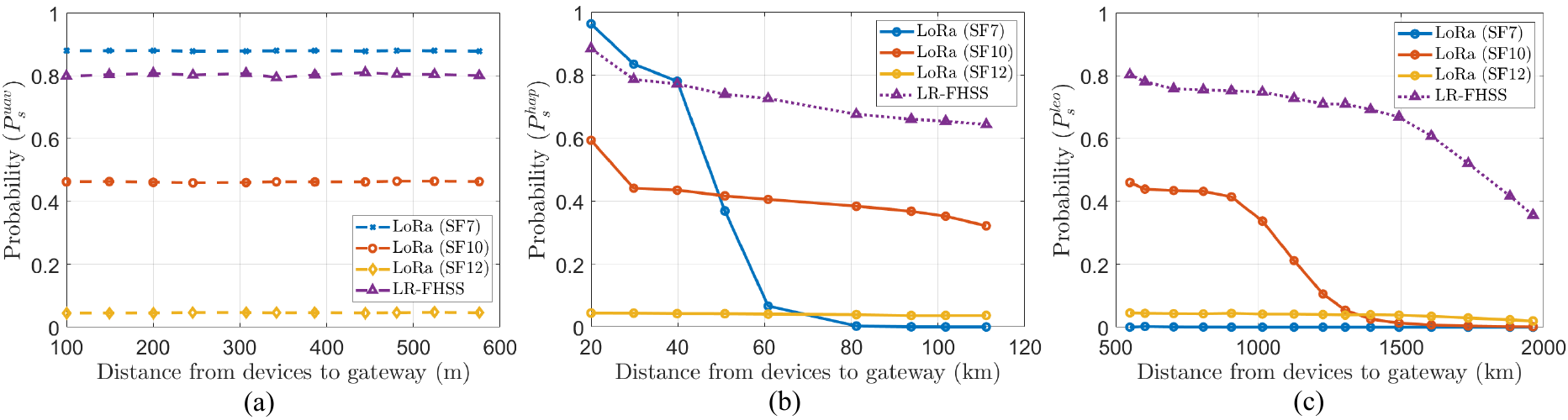}
    \caption{Average probability of packet delivery: (a) $P_s^{uav}$, (b) $P_s^{hap}$, and (c) $P_s^{leo}$ as functions of the distance from UDs to the UAV, HAP, and LEO satellite, respectively, for LoRa and LR-FHSS modulation schemes in a rural environment.}
    \label{fig_ModRes}
\end{figure*}

Fig.~\ref{fig_ModRes} depicts the average success probability of underground-to-NTN connectivity for LoRa and LR-FHSS modulation schemes as a function of the distance from UDs to the gateway in a rural environment, considering the UAV, HAP, and LEO satellite platforms. 

In Fig.~\ref{fig_ModRes}(a), the average $P_s^{uav}$ for LoRa SF7, SF10, SF12, and LR-FHSS is approximately $0.88$, $0.46$, $0.05$, and $0.80$, respectively, across all communication ranges under the UAV platform in a rural environment. The link budget for both LoRa and LR-FHSS is sufficient to cover the maximum distance, indicating $P_{snr}^{uav}=1$ for all ranges. Notably, LoRa SF7 exhibits the lowest collision probability due to its minimal time-on-air, leading to the highest $P_s$. Fig.~\ref{fig_ModRes}(b) illustrates that the average $P_s^{hap}$ decreases with increasing communication distance for the HAP platform in a rural environment. LoRa SF7 achieves the highest success probability up to $40$~km, beyond which LR-FHSS outperforms all LoRa SF configurations. This is primarily due to the limited link budget of LoRa SF7, which leads to a lower $P_{snr}^{hap}$ compared to LR-FHSS. With its higher link budget and greater resilience to interference, LR-FHSS provides superior performance at larger communication distances. Fig.~\ref{fig_ModRes}(c) shows that LR-FHSS delivers the best performance, maintaining $P_s^{leo} \geq 0.6$ from $550$~km to $1600$~km. Meanwhile, LoRa SF10 outperforms the other LoRa SF configurations up to $1300$~km, as SF10 provides more robust connectivity than SF7 and exhibits a lower co-SF interference probability compared to SF12. These findings highlight LR-FHSS as a promising wireless solution for underground-to-NTN connectivity in a rural environment. For LoRa modulation, SF7 is well-suited for short-range UAV communications, whereas SF10 is more appropriate for HAP and LEO satellite platforms. Accordingly, both LoRa SF10 and LR-FHSS are adopted to evaluate the performance of underground-to-NTN connectivity across various environments, UDs' numbers, burial depths, and VWC levels in Sections~\ref{SecResEnv},~\ref{SecResNum}, and~\ref{SecResUnderground}.

\subsection{Effects of Monitoring Environments} \label{SecResEnv}
\begin{figure*}[!t]
    \centering
    \includegraphics[width=6.9in]{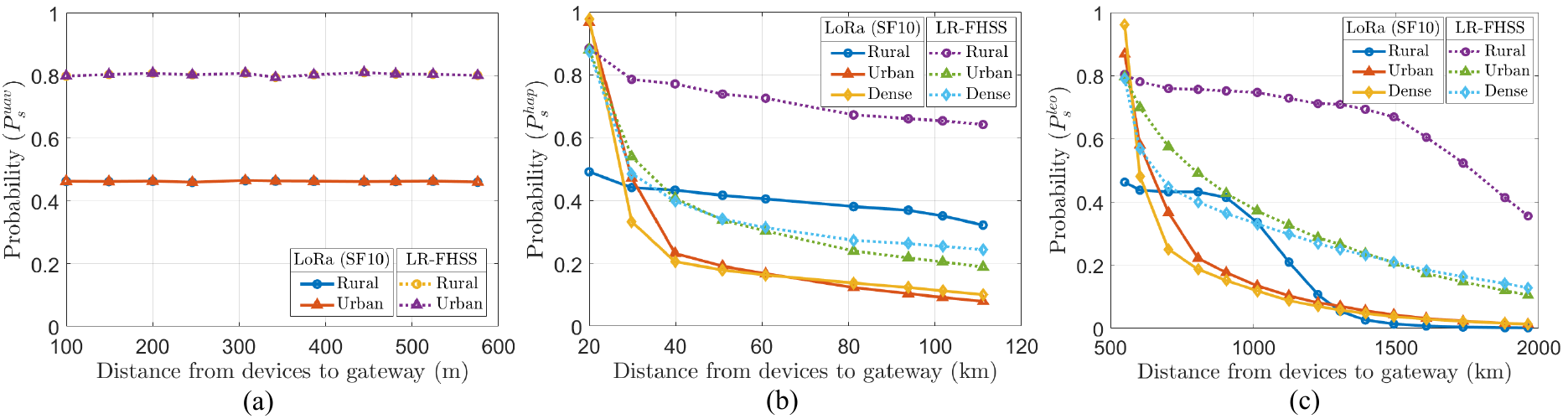}
    \caption{\textcolor{b}{Average probability of packet delivery: (a) $P_s^{uav}$, (b) $P_s^{hap}$, and (c) $P_s^{leo}$ as functions of the distance from UDs to the UAV, HAP, and LEO satellite, respectively, for LoRa SF10 and LR-FHSS considering various environments including rural, urban, and dense urban settings.}}
    \label{fig_ScenRes}
\end{figure*}

Fig.~\ref{fig_ScenRes} depicts the average packet delivery probability for LoRa SF10 and LR-FHSS as a function of the distance from UDs to the gateway, considering the UAV, HAP, and LEO satellite platforms in rural, urban, and dense urban environments.

Fig.~\ref{fig_ScenRes}(a) shows that the average $P_s^{uav}$ remains consistent across both rural and urban environments owing to the sufficient link budget and full LOS propagation conditions. Herein, LR-FHSS outperforms LoRa SF10, attaining a higher success probability of $P_s^{uav} = 0.80$ across all ranges, thanks to its superior resilience to interference. Fig.~\ref{fig_ScenRes}(b) represents that the average $P_{s}^{hap}$ in urban and dense urban environments is lower than that in rural settings due to the reduced LOS probability, increased clutter loss from buildings, and the effects of shadow fading. Notably, the average $P_s^{hap}$ in urban and dense urban environments exceeds that in rural environments at the highest elevation angle (i.e., $\theta=90^\circ$) due to the capture effect. The average $P_s^{hap}$ in urban environments is lower than that in a dense urban environment when the distance surpasses $80$~km, as a higher LOS probability is typically achievable in dense urban areas at lower elevation angles ($\theta \leq 15^\circ$). Furthermore, the average $P_s$ of LR-FHSS is higher than that of LoRa SF10 across all distances, except at $20$~km, where LoRa SF10 benefits from a higher probability of the capture effect occurring in urban and dense urban environments. Similarly, Fig.~\ref{fig_ScenRes}(c) indicates that the average $P_s^{leo}$ in urban and dense urban environments is significantly lower than that in rural environments. One can observe that the average $P_s^{leo}$ of LR-FHSS remains above $0.2$ at a distance of $1500$~km, even in challenging urban and dense urban environments. These findings highlight that the success probability is strongly affected by environmental conditions. Moreover, ensuring reliable underground-to-NTN connectivity remains a significant challenge.

\subsection{Effects of UDs' Numbers}\label{SecResNum}
\begin{table}[!t]
\caption{UDs' Densities for Various NTN Platforms Considering Different Numbers of UDs}
\centering
\renewcommand{\arraystretch}{1.2}
\label{DensityTab}
\begin{tabular}{|m{0.1\textwidth}<{\raggedright}|m{0.09\textwidth}<{\centering}|m{0.09\textwidth}<{\centering}|m{0.09\textwidth}<{\centering}|}
\hline
\textbf{Platform} & $N=10$k & $N=50$k & $N=100$k \\ \hline
\textbf{UAV} & 10k/km$^2$ & 50k/km$^2$ & 100k/km$^2$ \\
\textbf{HAP} & 0.3/km$^2$ & 1.3/km$^2$ & 2.7/km$^2$ \\
\textbf{LEO Satellite} & 0.0009/km$^2$ & 0.0045/km$^2$ & 0.009/km$^2$ \\
\hline
\end{tabular}
\end{table}

\begin{figure*}[!t]
    \centering
    \includegraphics[width=6.9in]{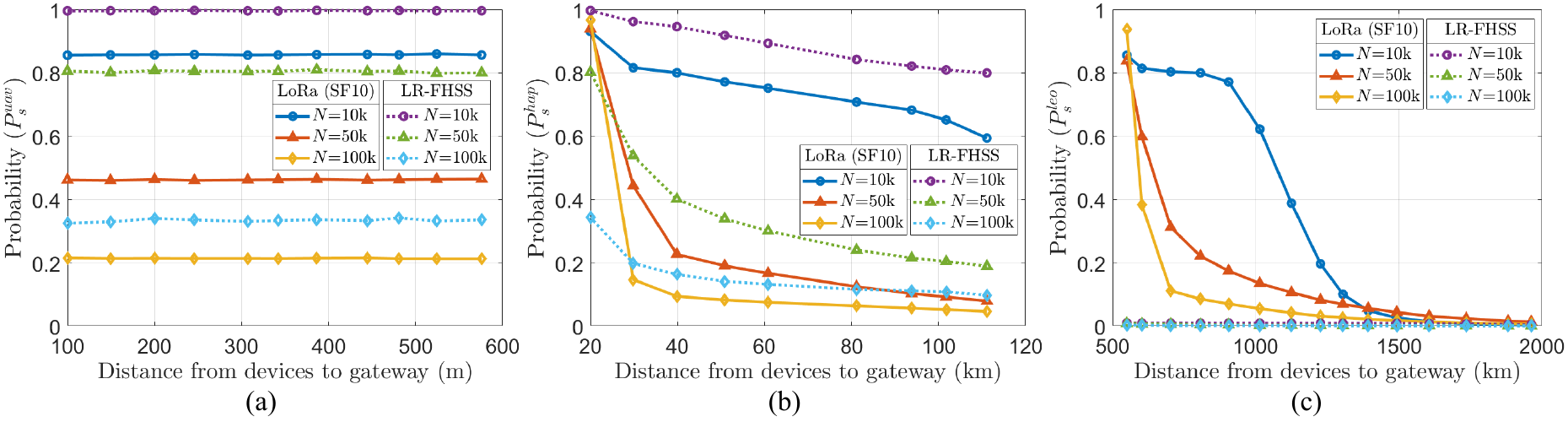}
    \caption{Average probability of packet delivery (a) $P_s^{uav}$, (b) $P_s^{hap}$, and (c) $P_s^{leo}$ as functions of the distance from UDs to the UAV, HAP, and LEO satellite, respectively, for LoRa SF10 and LR-FHSS under a rural environment considering different numbers of UDs.}
    \label{fig_NNRes}
\end{figure*}

Fig.~\ref{fig_NNRes} illustrates the average packet delivery probability for LoRa SF10 and LR-FHSS as a function of the distance from UDs to the gateway under varying numbers of UDs in a rural environment, considering the UAV, HAP, and LEO satellite platforms. The UD densities corresponding to different numbers of UDs for each NTN platform are summarized in Table~\ref{DensityTab}.

Fig.~\ref{fig_NNRes}(a) shows that the average $P_s^{uav}$ decreases as the number of UDs increases due to the higher collision probability. LR-FHSS consistently outperforms LoRa SF10 in terms of packet delivery ratio across all UDs' number cases. For instance, under $N=100$k, the average $P_s^{uav}$ for LR-FHSS is $0.12$ higher than that of LoRa SF10. Figs.~\ref{fig_NNRes}(b) and (c) depict the success probabilities of LoRa SF10 and LR-FHSS for the HAP and LEO satellite platforms, respectively, where the average $P_s^{hap}$ and $P_s^{leo}$ for LR-FHSS decline as the number of UDs increases. Notably, the average $P_s^{hap}$ and $P_s^{leo}$ for LoRa SF10 degrade as the number of UDs increases from $10$k to $100$k, except at $\theta=90^\circ$. This is attributed to the higher probability of capture effect occurrences, which are more pronounced with a larger number of UDs and the increased difference between individual UDs' channel budgets at the highest elevation angle. Furthermore, the average $P_s^{hap}$ and $P_s^{leo}$ for LR-FHSS exceed $0.8$ and $0.4$, respectively, even at the maximum distance with $N=10$k, demonstrating a significant improvement over LoRa SF10.

\subsection{Effects of Underground Parameters}\label{SecResUnderground}
\begin{figure}[!t]
    \centering
    \includegraphics[width=3.45in]{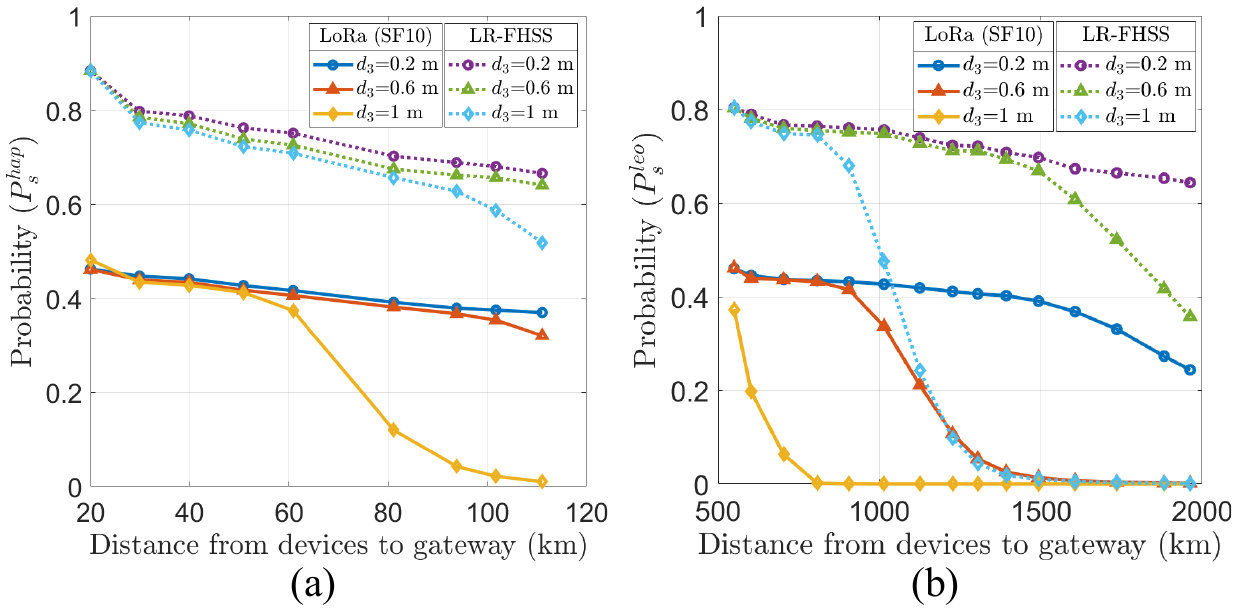}
    \caption{\textcolor{b}{Average probability of packet delivery: (a) $P_s^{hap}$ and (b) $P_s^{leo}$ as functions of the distance from UDs to the HAP and LEO satellite, respectively, for LoRa SF10 and LR-FHSS under a rural environment considering various burial depths.}}
    \label{fig_DepthRes}
\end{figure}

Pipelines are buried at varying depths depending on transportation requirements in underground pipeline monitoring applications. Therefore, it is crucial to evaluate the success probability of underground-to-NTN connectivity at different burial depths. Since the link budgets of LoRa SF10 and LR-FHSS are sufficient and $P_{sir}^{uav}$ is independent of burial depth, the average $P_{s}^{uav}$ remains consistent across the considered burial depths, as shown in Fig.~\ref{fig_ModRes}(a). Fig.~\ref{fig_DepthRes} illustrates the average packet delivery probability for LoRa SF10 and LR-FHSS as a function of the distance from UDs to the gateway at different burial depths in a rural environment, considering the HAP and LEO satellite platforms with a VWC of $11.19\%$.

Fig.~\ref{fig_DepthRes}(a) reveals that the average $P_{s}^{hap}$ degrades with increasing burial depth since a longer underground propagation path results in higher attenuation. For instance, at the maximum distance, the average $P_{s}^{hap}$ of LoRa SF10 drops from $0.37$ to $0.01$ when the burial depth increases from $0.2$~m to $1$~m. In contrast, the average $P_{s}^{hap}$ of LR-FHSS can achieve $0.52$ at the maximum distance, even at a depth of $d_3=1$~m. Fig.~\ref{fig_DepthRes}(b) indicates that the average $P_{s}^{leo}$ is also affected by burial depth, especially at larger communication distances. For instance, at a distance of $900$~km, the average $P_{s}^{leo}$ of LoRa SF10 decreases from $0.43$ to $0$ as the burial depth increases from $0.2$~m to $1.0$~m. Furthermore, the average $P_{s}^{leo}$ of LR-FHSS at $d_3=1$~m drops to $0$ at the maximum distance.

\begin{figure}[!t]
    \centering
    \includegraphics[width=3.45in]{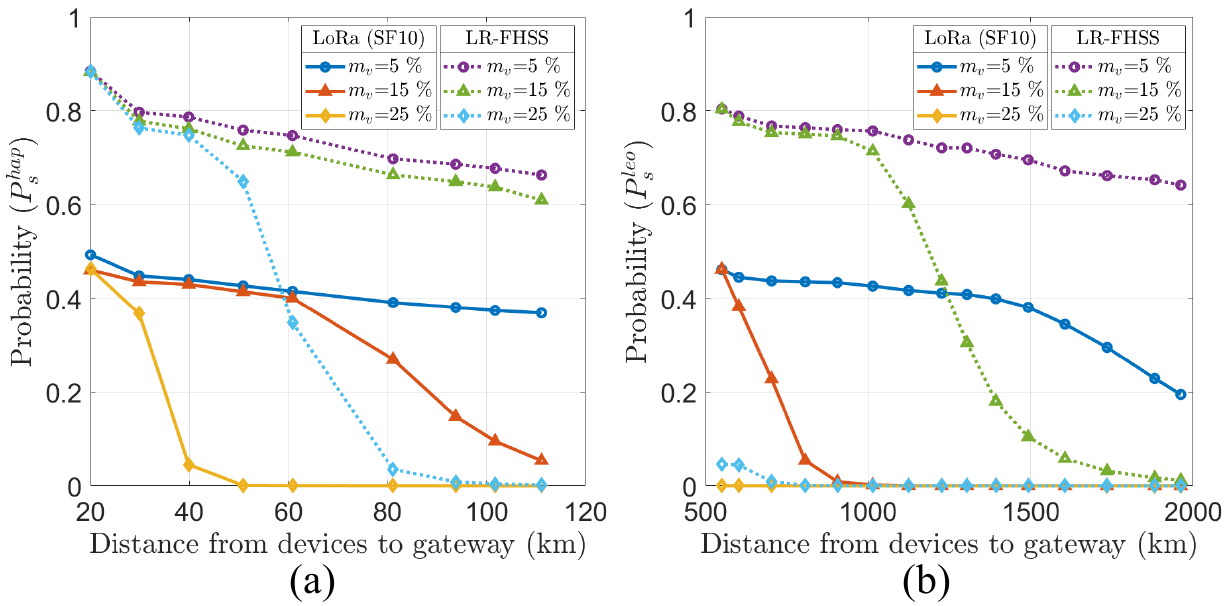}
    \caption{\textcolor{b}{Average probability of packet delivery: (a) $P_s^{hap}$ and (b) $P_s^{leo}$ as functions of the distance from UDs to the HAP and LEO satellite, respectively, for LoRa SF10 and LR-FHSS in a rural environment considering various VWC levels.}}
    \label{fig_VWCRes}
\end{figure}

Due to variations in VWC with precipitation, investigating the success probability of underground-to-NTN connectivity under different VWC conditions is crucial for underground pipeline monitoring applications. Since the link budgets of LoRa SF10 and LR-FHSS are sufficient for UAV communication and $P_{sir}^{uav}$ is independent of VWC, the average $P_{s}^{uav}$ remains constant across the given VWC levels, as shown in Fig.~\ref{fig_ModRes}(a). Fig.~\ref{fig_VWCRes} presents the average packet delivery probability for LoRa SF10 and LR-FHSS as a function of the distance from UDs to the gateway, under varying VWC levels in a rural environment, considering the HAP and LEO satellite platforms, with a fixed burial depth of $0.6$~m. 

Fig.~\ref{fig_VWCRes}(a) demonstrates that the average $P_s^{hap}$ deteriorates with increasing VWC, as higher VWC is a decisive factor in higher soil attenuation, which greatly influences path loss in soil. For instance, the average $P_s^{hap}$ of LoRa SF10 at a distance of $50$~km drops from $0.41$ to $0$ when the VWC increases from $5$\% to $25$\%. When $m_v\leq 15\%$, the average $P_s^{hap}$ of LR-FHSS remians above $0.61$, even at the maximum distance. Fig.~\ref{fig_VWCRes}(b) reveals that the success probability of underground-to-satellite connectivity is severely degraded under high-VWC conditions, with the average $P_{s}^{leo}$ for both LoRa SF10 and LR-FHSS approaching $0$ across all ranges when $m_v=25\%$. Therefore, UDs should pause uplink packet transmissions in high-VWC conditions to avoid unnecessary energy consumption.

\section{Conclusion} \label{ConclusionSec}
To extend the recently emerging NTN mMTC into the subterranean domain, this article proposed an underground-to-NTN connectivity framework and developed a simulator to evaluate the success probability of two LoRaWAN modulation schemes (i.e., LoRa and LR-FHSS) from UDs to various NTN platforms. Through extensive modeling of a realistic underground pipeline monitoring case, the numerical results demonstrated that the simulator could guide the selection of optimal modulation configurations for reliable communication across diverse deployment conditions and could be generalized to other underground monitoring scenarios by adjusting the channel model and environmental parameters. Note that the simulator did not account for the mobility of NTN platforms, and certain parameters require further site-specific calibration; both limitations will be addressed in our future work.

\bibliographystyle{IEEEtran}
\bibliography{ref}

\end{document}